# Pūioio: On-device Real-Time Smartphone-Based Automated Exercise Repetition Counting System

Adam Sinclair[1], Kayla Kautai[2], and Seyed Reza Shahamiri[3]

Department of Electrical, Computer, and Software Engineering

Faculty of Engineering,

The University of Auckland

Auckland, New Zealand

[1]kiaora@adamsinclair.kiwi, [2]kaylak106@hotmail.co.nz, [3]admin@rezanet.com

**Abstract**

Automated exercise repetition counting has applications across the physical fitness realm, from personal health to rehabilitation. Motivated by the ubiquity of mobile phones and the benefits of tracking physical activity, this study explored the feasibility of counting exercise repetitions in real-time, using only on-device inference, on smartphones. In this work, after providing an extensive overview of the state-of-the-art automatic exercise repetition counting methods, we introduce a deep learning based exercise repetition counting system for smartphones consisting of five components: (1) Pose estimation, (2) Thresholding, (3) Optical flow, (4) State machine, and (5) Counter. The system is then implemented via a cross-platform mobile application named Pūioio that uses only the smartphone camera to track repetitions in real time for three standard exercises: Squats, Push-ups, and Pull-ups. The proposed system was evaluated via a dataset of pre-recorded videos of individuals exercising as well as testing by subjects exercising in real time. Evaluation results indicated the system was 98.89% accurate in real-world tests and up to 98.85% when evaluated via the pre-recorded dataset. This makes it an effective, low-cost, and convenient alternative to existing solutions since the proposed system has minimal hardware requirements without requiring any wearable or specific sensors or network connectivity.

## 1 INTRODUCTION

Regular exercise or "workouts" is important for maintaining personal health and well-being [1]. Workouts typically involve repeating one or a collection of exercises to maintain or improve physical fitness. Therefore, tracking the number of exercise repetitions completed is advantageous first to calibrate the current workout to the appropriate difficulty and, secondly, to measure progress between workouts over time. Additionally, adjusting the number of reps can help achieve various goals, such as improving fitness and health, increasing functional strength, or building definition and bulk [2].

However, during periods of exertion or fatigue common during strenuous activity, maintaining a mental count of the number of exercises completed can present a challenge. Adopting an application that requires manual input for workout logging is a widespread but poor solution to this problem. Previous studies applied deep learning to accomplish Automated Exercise Repetition Counting (AERC) and reported high accuracies. However, there was no successful attempt to create an on-device, real-time Computer Vision (CV) based AERC system. This is because the existing approaches are mostly unsuitable for real-time applications on computationally less capable devices such as smartphones without dedicated deep learning processing capabilities such as graphical, neural, or tensor processing units [3]. An alternate solution is to conduct inference online on servers that would require stable and fast network connections, with which the network lag may still render any real-time usage scenario impractical. Other studies also proposed solutions that mainly require physical sensors or wearable devices to be worn by the person exercising, which incur extra costs or suffer from lower accuracies.

To save the cost and inconvenience of wearing such devices during exercise and to overcome the limited processing capability of mobile devices, this study investigated a system for the automatic, real-time counting and recording of exercise repetitions using only a single smartphone camera recording the person exercising. In particular, we investigated how to design an architecture that performs AERC using CV from a single video source and how the proposed architecture could be deployed as a mobile application that performs AERC offline and only using the smartphone camera without the need for any extra wearable and sensor, or network connectivity.

In this paper, after providing a comprehensive overview of different AERC methods and their limitations, we propose a lightweight AERC approach that integrates pose estimation and optical flow and applies thresholding and state machines to enable real-time exercise repetition counting on typical smartphones. To verify the performance of the application, two experiments were conducted, including real-life testing of the system.

This paper is organized as follows. Section 2 provides an overview of the existing methods that could be used for AERC and compares them. Section 3 presents the proposed Automatic Exercise Repetition Counter, its architecture, and its components. Section 4 introduces Pūioio, the application that deploys the proposed system on smartphones. The evaluation methods and experimental results are discussed in Section 5. Finally, Section 6 discusses the results, limitations of the proposed system, and directions for future studies and concludes the paper.

## 2   RELATED WORK

Approaches that could be used to count exercise repetition can be divided into two general categories: sensor-based or computer vision-based. While the former has been widely used in the literature due to wearables becoming widespread consumer gadgets for continuously tracking physical activities in close to real-time, computer vision approaches are more recent and have shown promising results. In this section, we explain and review both approaches.

## 2.1    Sensor-based approaches

As the most common approaches for AERC, sensor-based methods use a device to detect an input from a physical environment to generate a signal representing the physical phenomenon. We can classify these methods into four categories, shown in Figure 1, based on how they sense movement and capture data. The following sections provide an overview of the four categories regarding their methods and accuracy (%) – the degree to which the observed results conform to the correct number of reps performed.

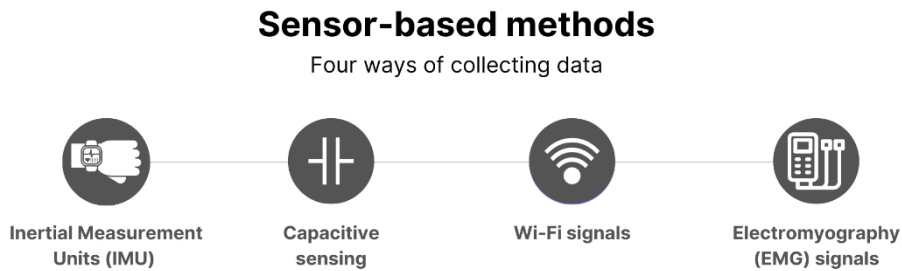

**Figure 1.** Sensor-based categories

### 2.1.1    Inertial Measurement Units (IMU) AERCs

IMUs primarily employ inertial sensors, such as accelerometers [4] and gyroscopes [5], [6], positioned on the user's body. Among the most popular IMU approaches is Peak Detection [6], which pre-processes the sensor data into a 1D signal to count the number of strong peaks as reps; this approach was adopted by [4], [5]. Nonetheless, a limitation of this approach is depending on the exercise and individual's form, the data can have multiple peaks per repetition, fluctuating in the repetition rate or causing variations in peak shape and amplitude that may lead to inaccurate counting. Previous studies reported generating false positives or negatives for all exercises, for example, studies conducted in [4], [6]. To counteract this, Das et al. [5] used amplitude thresholding for each exercise to reduce any inaccurate peaks caused by noise. This approach led to the most accurate IMU method, as the authors reported their approach detected valid repetitions 99.4% of the time from the accelerometer signal collected using a smartwatch.

Among other IMU approaches are the use of the Viterbi algorithm [7] that relies on Hidden Markov Models to count reps (90% accuracy reported), Artificial Neural Networks where raw inertial sensor data were inputted into a neural network to output a binary classification of either zero or one, in which a series of 1's indicates a repetition (best accuracy of 73.5% reported), utilizing feature extraction and selection on the accelerometer and gyroscope data [8] (93% accuracy); and Dynamic Time Warping [9] to determine the similarity of the actual rep and the observed rep pattern (61% accuracy reported).

### 2.1.2    Wi-Fi-Signal AERCs

The human body can reflect Wi-Fi signals to a receiver when a Wi-Fi transmitter and receiver are present. Any body movement causes the reflected Wi-Fi signal to undergo a Doppler frequency shift. Recording this shift allows for feature extraction resembling bodyweight activities to quantify the number of repetitions. Previous studies leveraged this feature and designed AERC methods with a pair of off-the-shelf Wi-Fi devices [10]–[12]. A notable work is [10] that achieved 99% accuracy using an impulse-based method to segment and count the number of repetitions. However, the work is limited as it is not adaptable to changes in the environment, such as moving furniture, or the Wi-Fi receiver or transmitter, which may be inconvenient for users, especially those who exercise at gyms.

### 2.1.3    Capacitive Sensing AERCs

Another less frequently explored method is collecting data from capacitive coupling-based sensors. An example is [13] in which a peak detection algorithm was applied to capacitance values recorded by a specially designed wrist wearable. As a result, the approach achieved a counting accuracy of 91%. However, this approach is still hindered by the peak detection algorithm counting false peaks from noise regardless of the initial filtering steps.

### 2.1.4    AERCs using Electromyography (EMG) Signals

The use of EMG signals was studied in [14], where EMG signals were recorded using a small, wireless electromyograph placed close to the concerned muscles. The detected signals were then used to find the mean frequency of the amplitude spectrum, in which patterns imply repeated movements. Although this study attained an accuracy of 97.1%, this approach is limited since the electrodes require direct contact with the user's muscle or skin with minimal impedance to produce a reliable signal. Therefore, recording EMG signals could be impractical for workouts as the user precipitates and moves excessively. Likewise, placing the electrodes requires pinpoint accuracy, likely leading to human error in real-world applications.

## 2.2    Computer Vision-based Approaches

There is relatively limited literature on automatic exercise repetition counting without wearable devices or external resources further than a common smartphone. Of the literature which did explore this, their methods can be classified into three categories: (1) those which employed Pose Estimation, (2) those which employed Optical Flow, or (3) a repetition counting model. In this section, we review these approaches.

### 1.1.1 Pose Estimation methods

A crucial component of a computer vision-based counting pipeline is the machine's ability to depict humans and their interactions visually. Hence, pose estimation has been most commonly applied to determine a person or object's stance from digital visual input such as an image or video. It involves a real-time network that detects human poses from an input video feed and extracts a 2D or 3D skeleton of key points from these detected poses [15]. One of two approaches is then taken: (1) using the relative angle and distance between the estimated key points to track poses within frames, and hence count exercises, or (2) classifying video frames into "Key Actions" representing a certain important part of an exercise repetition, using an internal state machine to count repetitions. The human pose estimator's effectiveness has increased due to the development and advances in Deep Neural Networks (DNNs).

OpenPose [16] is a real-time, multi-person 2D pose estimator and has been among the most widely used pose estimation approach in the literature. An example output of OpenPose can be seen in Figure 2. Among other pose estimation models is the Multi-Stage Network for Human Pose Estimation (MSPN) model [17]. While OpenPose employs a bottom-up approach, first predicting joints and then associating them with individuals, MSPN uses a top-down approach that first locates the person and then applies a single-person pose estimator to predict key point locations. An example of applying MSPN is [18] in which the authors indicated its high accuracy and stability. Among other notable deep learning-based pose estimators are Mask R-CNN [19] and AlphaPose [20].

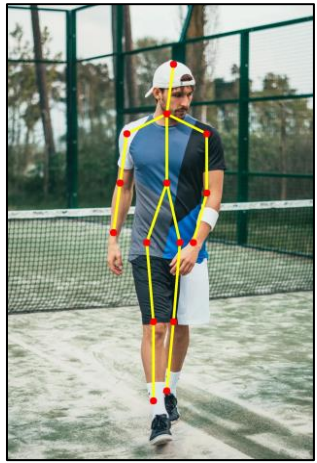

**Figure 2.** The output of the OpenPose model overlayed on an input image [21]

Even though the methods listed above are state-of-the-art, they demand significant computing power and are based on complex models, which makes them less suitable for real-time performance on mobile devices. Therefore, "lightweight" versions have been optimized to install on devices with limited hardware, for example, Lightweight OpenPose [22] and MoveNet [23]. MoveNet is a high-speed and accurate model that detects 17 key body points of an individual, whereas Lightweight OpenPose detects 18 key points for multiple humans. A recent study comparing OpenPose, MoveNet, and PoseNet concluded that MoveNet was the fastest, but PoseNet delivered better accuracies [24].

Unlike 2D pose estimators mentioned above, 3D approaches determine the precise spatial positioning of a human or object by transforming a 2D image into a 3D object. This method is more challenging than 2D methods as it requires estimating an additional $z$-dimension. Although many 3D methods like DensePose [25] have proven to be precise and efficient, the algorithms are incompatible with low computing powered devices. Compact models such as BlazePose [26] were introduced, but their applications may be limited. For example, BlazePose detects a single person whose face must be visible to identify a pose.

### 2.2.1 Optical Flow Methods

Optical flow refers to the apparent movement of pixels between contiguous frames resulting from an object's actual movement in the environment with respect to the camera. It works under the assumption that pixel intensities of an object do not change between consecutive frames, and that neighboring pixels have similar motions [27]. When this is performed on a subsampled series of features chosen from the image, it is referred to as Sparse Optical Flow (SOF) [28]. When it is instead performed on all pixels, it is referred to as Dense Optical Flow (DOF) [29]. Intuitively, DOF is more computationally expensive than SOF. An example output of Optical Flow applied to a push-up exercise is shown in Figure 3.

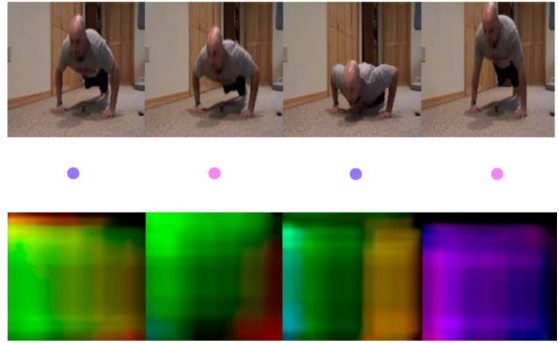

**Figure 3.** Dense optical flow output for a push-up exercise repetition [30]

### 2.2.2 Repetition Counting Models

A RepNet model was introduced in [31] claiming to be able to count any repeating action rather than solely workout repetitions. The detection architecture included three primary components: the learned encoder, the period predictor, and a temporal self-similarity layer separating them. In this approach, video frames are fed to the encoder to create embeddings. Then, these embeddings are collapsed in the spatial dimension, removing the need for explicit region of interest tracking. Using these embeddings, the pairwise similarities between pairs of embeddings are used to produce a self-similarity matrix as the key predictive feature for the model. This matrix is fed into the period predictor to produce a period length estimate and a periodicity score. The former is the predicted rate at which repetitions occur, while the latter indicates whether the given frame was within a periodic portion of the video. An aggregation of the period length produces a repetition count. Videos can be sampled at different frame rates to predict larger period lengths.

### 2.2.3 Performance Comparison and Limitations of the Existing CV-based AERCs

Table 1 summarizes and compares the aforementioned computer vision-based models used for AERC. It should be noted that a truly fair comparison of these methods is challenging due to the varying datasets and evaluation methodologies employed, and any differences should be considered with this caveat. Concerning the evaluation methods used, Off-by-One (OBO) Accuracy is the model's accuracy across multiple tests, where an inference is successful if the absolute difference between the predicted count and actual count is less than or equal to one. MAE refers to Mean Absolute Error as the sum of absolute errors divided by the sample size.

**Table 1.** Comparison of Computer Vision-Based Models for Automated Exercise Counting

| Methodology Type | Reference | Methodology | Evaluation Method | Score | Evaluation Method 2 | Score 2 |
|---|---|---|---|---|---|---|
| *Pose Estimation* | [15] | Pose Estimation with Thresholding | OBO accuracy | 79%1 | MAE | 0.1472 |
| | [32] | Pose Estimation with Heuristic and DTW | F1 Score | 0.86 | - | - |
| | [33] | Pose Estimation with Thresholding | Count accuracy | 94-95% | MAE | 0.06 |
| | [34] | Key Moment Classification | Count accuracy | 97.1% | MAE | 0.029 |
| | [35] | Key Moment Classification | Count accuracy | 91.6% | - | - |
| | [36] | MSPN Model with Thresholding | OBO accuracy | 99% | MAE | 0.004 |
| *Optical Flow* | [28] | Lucas-Kanade Sparse Optical Flow | Count accuracy | 96% | - | - |
| | [29] | Temporal Gunnar Farneback Dense Optical Flow | OBO accuracy | 79.6% | - | - |
| *Other* | [31] | RepNet Model | - | - | MAE | 0.3641 |

Pose estimation with thresholding is a methodology weighted towards speed rather than accuracy. As visible in Table 1, count accuracies achieved for this method range between 80-95%. However, the inference time required to achieve this is significantly less, considering the pipeline consists of only a pose estimation model and analytical methods of thresholding rather than a multimodal inference architecture. As the pose estimation model may also be transferred from other domains, the amount of data required for this methodology is significantly less than those needing to train a CNN. In fact, in theory, if the thresholds are determined by a trained professional, only a test set is required so that model accuracy may be ascertained.

Comparatively, key moment classification methodologies achieved better accuracies. However, such approaches require multiple CNNs to be trained on a per-exercise basis. These CNNs necessitate a much greater volume of training data to achieve accurate count predictions, posing data collection challenges. Furthermore, as the key moment classification is typically combined with a pose estimation model, this requires a multimodal architecture that is more computationally expensive and hence has slower inference times. As such, they are not suitable for computationally limited mobile and real-time environments.

---

1 As reported by [36].

2 As reported by [36].

RepNet is a promising model, with the advantage of improved generalization compared with the other methodologies, as a single RepNet model may count any exercise repetition. However, the model size results in relatively slow inference times, making it less applicable for real-time performance. The model size and complexity also represent significant challenges when deploying the model on mobile platforms. Given the model's creation in Tensorflow, the model's architecture necessitates operations outside of the standard TensorFlow Lite runtime [37], requiring experimental, extended binaries to operate successfully, further adding to any mobile deployment challenge.

A common issue mentioned across pose estimation studies is the limitation caused by the lack of availability of large datasets. Likewise, noticing an imbalance between key phases in its dataset (for example, only a few samples of the top of squat, many examples of moving down), a study conducted in [35] performed key pose augmentation to achieve balanced data. Furthermore, several studies noted that different environments and physiologies could negatively impact results, such as the difference between male and female stature and joint locations. Normalization compared to the torso height should help to mitigate this issue but may still present a challenge for participants with outlier statures.

From reviewing [28], [29], Optical Flow alone seems unable to deal with more complex exercises. Instead, it appears best suited for exercises with simple, large movements. Accordingly, exercise-specific models will likely achieve the best results, especially when including exercises with compound movements or greater complexities. Furthermore, the use of pose estimation may facilitate basic form correction, as explored in [15], [32], [34].

## 3  The Proposed Automatic Exercise Repetition Counter

The proposed AERC is composed of five components, as shown in Figure 4, and operates on a frame-by-frame basis by processing two consecutive frames at each point in time. With each incoming frame, pose detection (1) is first performed before thresholding (2) is applied to determine the exercise's "key moment". Alternatively, Optical Flow (3) is performed if the pose detection model has insufficient confidence. If the pose detection model remains insufficiently confident for longer than a predetermined length, initially 1.5s, state transitions are paused till confidence is regained. Then, either pose detection or optical flow's output is fed to the state machine (4) that determines the current "movement phase". Three movement phases are identified for basic exercises like squats, push-ups, and pull-ups: top, intermediate, and bottom. As each frame is processed independently, no distinction can be made in either thresholding or optical flow between the descending and ascending motions. Accordingly, these are deduced from context [38]. The movement phase determined by thresholding or optical flow is then passed into an exercise-specific state machine. Once one cycle of state machine transitions has been completed, the repetition count is incremented by one (5). In the following, each component is further explained.

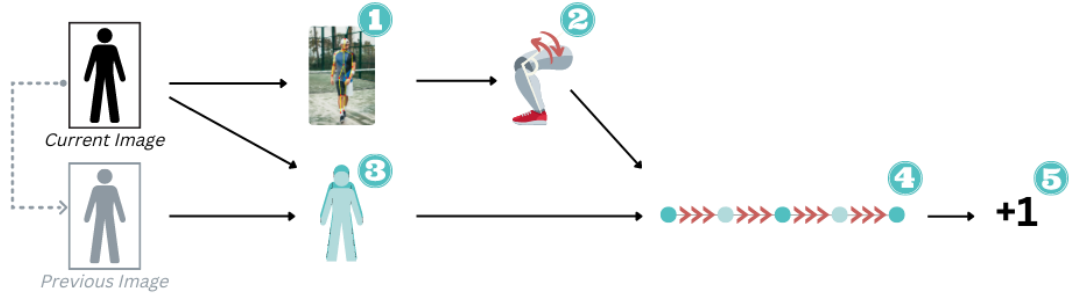

**Figure 4.** Illustration of the overall system, comprising five components: (1) Pose Detection, (2) Thresholding, (3) Optical Flow, (4) State Machine, and (5) Counter

## 3.1 Pose Detection

The pose detection component is based on Google's BlazePose model [26], which uses a CNN to estimate human pose. This CNN architecture is selected since it is lightweight and capable of performing inference in real-time, on-device. Pre-trained and provided as part of Google's Machine Learning kit, this component takes an image tensor as input and outputs 33-point detections corresponding to features on the detected body, as shown in Figure 5. These point detections contain cartesian coordinates and the estimated probability of the point being in the image.

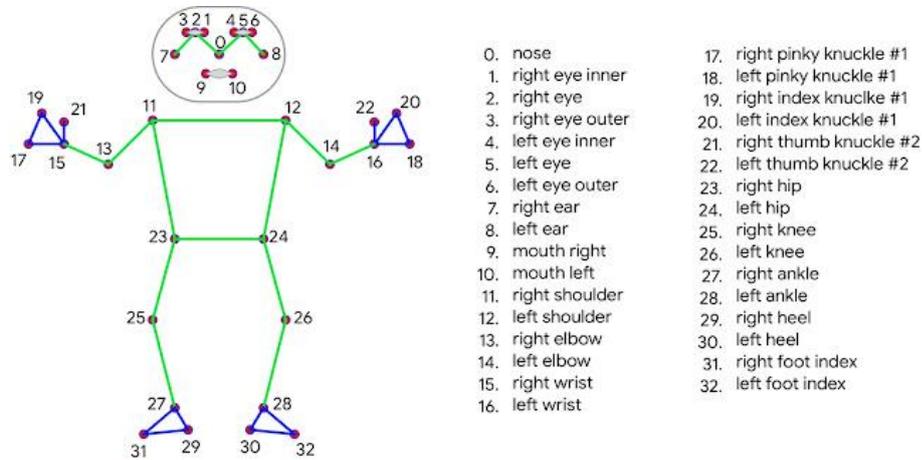

**Figure 5.** Pose Detector's 33 key point topology [26]

Once the detections have been calculated, the key points for a given exercise are extracted. In the example of a Squat, these key points include the right and left legs, specifically points 24, 26, and 28 on the right side, and 23, 25, and 27 on the left side of Figure 5. These key points are then checked for confidence; if the likelihood of any key point being in the image is less than a pre-tuned threshold, the pose detection is discarded, and optical flow is instead used. If the confidence of all points is sufficient, the detections undergo thresholding.

## 3.2    Thresholding

The thresholding component calculates the current movement phase as either 'Top', 'Ascending', 'Descending,' or 'Bottom' by using the key points predicted from the pose detection component. These phases represent the fundamental poses required to count a repetition of a particular exercise and are determined using exercise-specific angles.

From the raw pose detection points, the internal joint angles are calculated. The internal joint angle is the internal angle made by connecting three constituent joint coordinates, as shown in Figure 6.

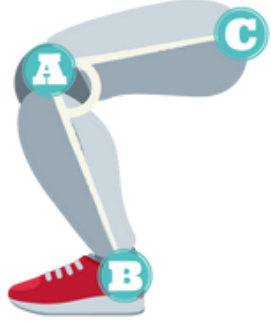

**Figure 6.** Key joint angle calculation

The cosine rule (Eq. 1) determines the cosine of the angle between two vectors given those vectors. In Figure 6, two vectors are formed between the internal point (A) and the two external points (B, C), respectively. Accordingly, the internal angle is calculable using a rearrangement of this cosine rule (Eq. 2).

$$\cos \theta = \frac{\vec{u} \cdot \vec{v}}{(\|\vec{u}\| \cdot \|\vec{v}\|)} \qquad \text{Eq. 1}$$

$$\theta = \cos^{-1} \frac{\vec{u} \cdot \vec{v}}{(\|\vec{u}\| \cdot \|\vec{v}\|)} \qquad \text{Eq. 2}$$

where $u$ and $v$ are the two connected vectors, and $\theta$ is the angle between them. With the internal joint angle determined, movement phases are extracted using predefined thresholds. Table 2 is the conversion table for this, while the predefined thresholds are provided in

**Table 3**. Pull-ups are configured oppositely in Table 2 due to their ascending rather than descending second phase.

**Table 2.** Current Angle to Movement Phase Per Exercise

| Exercise | Current Movement Phase (*Given Current Angle $\theta$*) | | |
|---|---|---|---|
| | $\theta$ < *Lower Threshold* | *Lower Threshold* ≤ $\theta$ ≤ *Upper Threshold* | $\theta$ > *Upper Threshold* |
| *Squats* | Bottom | Intermediate | Top |
| *Push-ups* | Bottom | Intermediate | Top |
| *Pull-ups* | Top | Intermediate | Bottom |

**Table 3.** Exercise Thresholds

| Exercise | Lower Threshold Angle | Upper Threshold Angle |
|----------|----------------------:|----------------------:|
| *Squats* | 130° | 150° |
| *Push-ups* | 130° | 150° |
| *Pull-ups* | 100° | 150° |

To increase robustness to noise, a sliding window approach is also used to determine the output movement phase for both thresholding and optical flow. That is, the current movement phase is deemed to be the average movement phase over a sliding window of movement phases for previous frames, including the current frame. An example sliding window of size five is shown over one movement phase addition in Figure 7, in which "A" stands for "Ascending" and "T" for "Top". The size of the window is a configurable parameter discussed in the following sections.

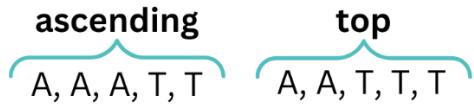

**Figure 7.** Example sliding window of size five

## 3.3 Optical Flow

If the confidence of any pose detection key point is below the predefined threshold, Optical Flow is instead used to determine the movement phase. Optical Flow is adopted as a secondary input due to its lesser specificity and accuracy when compared to pose estimation. Optical Flow is applied to determine motion between frames, aka "flow", as illustrated in Figure 8. The "movement" or "flow" is detected by tracking adequate points between two frames – in our case, the current and previous frames in **Figure 4**. Once the key points are averaged, the overall movement is determined with a general direction of 'Up', 'Down', or 'Stationary'.

Our Optical Flow component is based on Lucas-Kanade Optical Flow, as provided in [39]. Here, tracking points are selected every few frames, with these points being tracked between subsequent frames. As previously explained, Optical Flow is best used for tracking large movements within the image, with the necessary assumption that the exercise is the dominant movement in the frame.

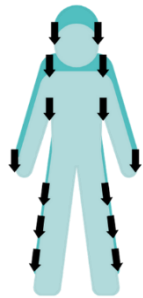

**Figure 8.** Sparse Optical Flow

Only those points with flow above a threshold are included to avoid including tracking points not on the person (such as a doorway behind the subject in the frame). From the $X$ and $Y$ Cartesian displacement of each of these points, the movement direction is extracted from the polarity of the flow: left, right, or stationary in the $X$ axis; up, down, or stationary in the $Y$ axis. For a given exercise, only the flow parallel to the movement is considered. For example, in the case of a squat, only the $Y$ flow is included. This is because squatting is a vertical exercise with no lateral movement; hence, any lateral flow is considered noise. Flow is calculated as the main movement direction of these active, valid points.

To increase robustness to noise, especially given the imprecise nature of optical flow, a sliding window approach is used to determine the current optical flow itself. That is, the current optical flow is deemed to be the average optical flow over the sliding window. Given the current (average) flow, previous (average) flow, and previous movement phase, the current output phase is calculated with the conversion shown for a vertical exercise in Table 4.

**Table 4.** Optical Flow to Movement Phase Conversion for a Vertical Exercise

| Current Optical Flow | Previous Optical Flow | Previous Movement Phase | Output Movement Phase |
|---|---|---|---|
| Up | Stationary | Bottom | Intermediate |
| Up | Up | - | Intermediate |
| Down | Stationary | Top | Intermediate |
| Down | Down | - | Intermediate |
| Stationary | Stationary | Bottom | Bottom |
| Stationary | Down | Intermediate | Bottom |
| Stationary | Stationary | Top | Top |
| Stationary | Up | Intermediate | Top |

## 3.4 State Machine and Counter

Once the predicted movement is determined by either the pose estimator or optical flow component, the state machine component is engaged. The state machine takes as input the current movement phase as determined by thresholding or optical flow, using this input to determine the current exercise "moment". For a given exercise, states are predefined under consultation from professionals. In the example of a squat, four key moments are identified: (1) standing, (2) descending, (3) squat, and (4) ascending, as shown in Figure 9.

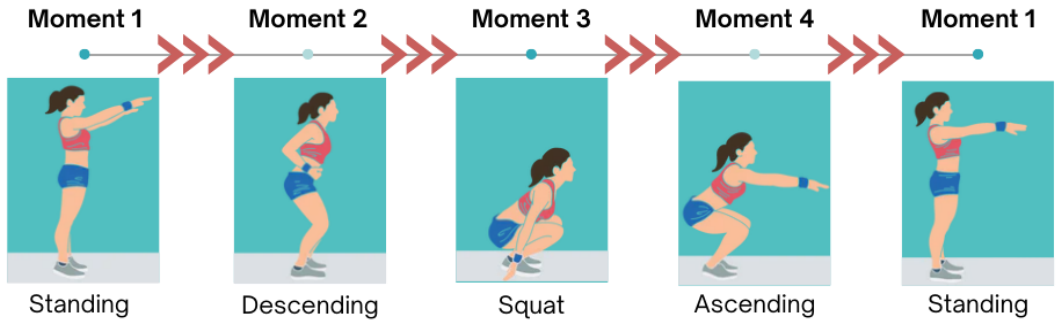

**Figure 9.** Key moments in a squat exercise

As a state machine, transitions are determined by both the input (the current movement phase) and the current state (the current exercise moment state). Figure 10 shows the state machine diagram for a basic vertical exercise. Any input that does not match the transition condition causes no change in state. Note it is also not possible to transition backward between states. This simplification prevents erroneous transitions due to noise and improves accuracy. It also encourages participants always to complete a full repetition. In the squat example, a person who does not squat deep enough – in that they do not satisfy the threshold to move into the 'Bottom' phase – cannot transition to other phases until the 'Bottom' phase is fulfilled. However, if the system fails to detect a phase change, the user would be required to fulfil a complete sequence again. Once a cycle of states has occurred, the exercise repetition counter component is incremented by one.

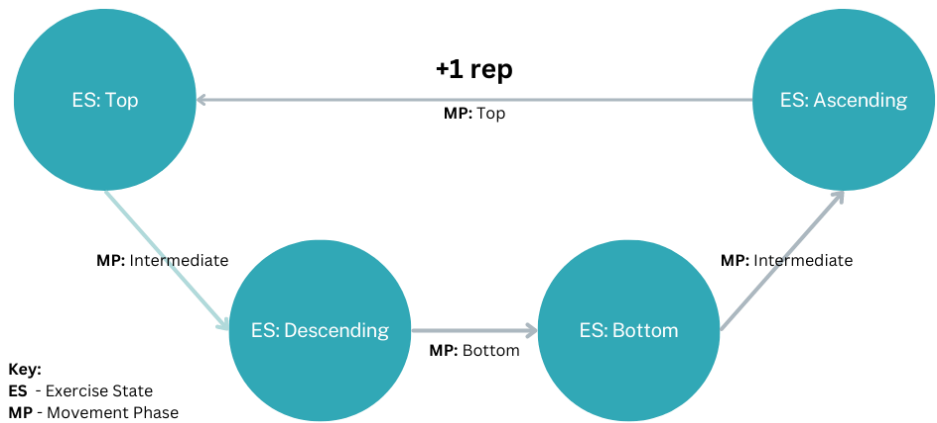

**Figure 10.** Vertical exercise state machine diagram

## 3.5  Hyperparameter Tuning

Given the nature of the methodology, many elements needed to be tuned to achieve optimal results. A summary of these hyperparameters and their empirically determined value can be seen in Table 5.

**Table 5.** Hyperparameter Variables and Values

| Hyperparameter | Empirically Determined Value |
|---|---|
| Application pause timeout | 1.5s |
| Exercise thresholds | *See Table* **3** |
| Pose detection in-frame confidence threshold | 0.75 |
| Sliding window size, thresholding | 1 |
| Sliding window size, optical flow | 3 |
| Optical flow movement threshold | 5 |

## 4 MOBILE APPLICATION DEVELOPMENT

The proposed AERC is implemented and deployed into a mobile application named Pūioio (a te reo Māori word meaning "strength") that is available on the Google Play Store [40] and Apple App Store [41]. The app allows users to track their repetitions of either Squats, Push-ups, or Pull-ups solely using their smartphone's front-facing camera and works without access to the Internet and in real-time. This section explains the application.

Pūioio's user interface is shown in Figure 11. To start counting reps for an exercise, a user can navigate to the *Quick Start* (Figure 11 (a)) tab using the bottom navigation bar. On this page, the user is required to select an exercise and their desired number of repetitions. Specifically, a user can swipe left or right on the image carousel to choose from the three exercises. Furthermore, they can click the plus or minus icon buttons to add and remove reps. Pressing the 'Begin' button starts the AERC process. Before starting the repetition counting, the app instructs the user on how to set up their phone and complete the exercise, as seen in Figure 11 (b). Clicking the button "I'm ready!", initiates a ten-second countdown to provide the user sufficient time to initiate their starting position. During the countdown sequence, the phone's front-facing camera is launched.

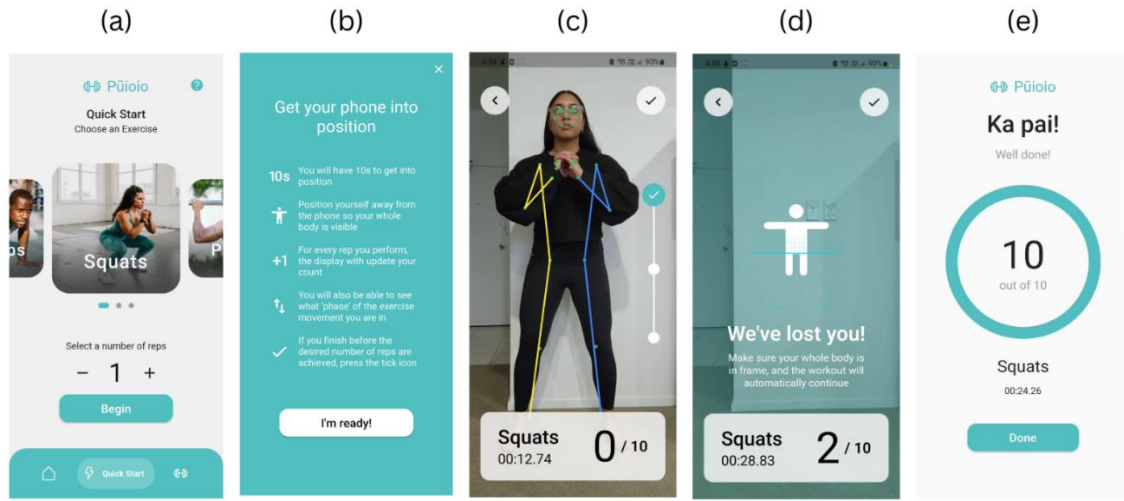

**Figure 11.** Pūioio app user interface (a) Quick Start tab; (b) Get Started screen; (c); Automated Counting screen; (d) User not detected warning; (e) Results screen

Once the countdown reaches one, the Automated Counter (AC) module begins the counting process (Figure 11 (c)). During the exercise, the user can view the following:

A "skeleton" is painted over the camera input with a select number of key points generated from the pose estimation.

A phase tracker aligned to the right to inform the user when they have reached an exercise's full extension or depth. Each stage on the tracker, from top to bottom, corresponds relatively to the exercise phases top, ascending/descending, and bottom.

An information panel displaying the exercise name, a stopwatch running until the exercise is complete, and the repetition count generated from the model.

The app automatically navigates to the Results page (Figure 11 (e)) once the model's predicted count equals the desired reps. However, if the model cannot confidently detect the pose of the user, the counter will be paused and display the message shown in Figure 11 (d) to the user. Once the user is fully in frame, the model will automatically continue to count repetitions.

## 5  EVALUATION AND RESULTS

To evaluate the proposed AERC, we conducted two sets of experiments. The first evaluation experiment was conducted using a custom dataset of pre-recorded videos showing individuals performing exercises. The second set of experiments was conducted via Pūioio in real-life scenarios by two participants. Counting accuracy and Off-by-one (OBO) accuracy were measured during each experiment.

### 5.1  Evaluation via Custom Dataset

To the best of our knowledge, no publicly accessible datasets include video recordings of participants completing repetitions of various exercises with a repetition count. Nevertheless, the free video-sharing platform, YouTube, hosts millions of fitness videos uploaded by various users. As a result, we curated a video-based dataset of individuals performing three standard exercises: Squat, Push-up, and Pull-up. After finding appropriate videos, the footage was trimmed to the section where the individual completed the set of repetitions. Additionally, the video was cropped to the most common smartphone aspect ratio (16:9). Lastly, we manually counted the number of repetitions in each video and labeled each sample. In total, 30 videos for each exercise (an average of 9.5 minutes per exercise) were collated and consisted of various genders, ethnicities, body types, lighting, and angles. Table 6 provides more information about the dataset.

**Table 6.** Overview of the Custom Dataset

| Exercise | # Videos | Total duration |
|----------|----------|----------------|
| Squats   | 30       | 11 m 1 s       |
| Push-ups | 30       | 8 m 44 s       |
| Pull-ups | 30       | 8 m 46 s       |

Then, each video of the dataset was converted into a sequence of images and fed frame-by-frame into the proposed AERC. After the tests were successfully executed, a debug output was generated by the AERC, recording the number of repetitions counted for each video of the exercise being tested. A summary of the results can be seen in Table 7. It is pertinent to note that the mean metrics presented in Table 7 were the accuracies averaged across each video as a standalone experiment. That is, each video was considered as a standalone trial in which accuracy and OBO were measured. For each exercise type, 30 accuracy and OBO metrics were measured, with their mean value presented in Table 7. In contrast, the absolute accuracies provided are the ratio of the overall number of reps predicted across all exercise videos (per exercise type) to the total number of actual reps. Accuracy was calculated as follows:

$$Error\ Rate = \frac{abs(\ \#Actual\ Reps - \#Predicted\ Reps)}{\#Actual\ Reps} \qquad \text{Eq. 3}$$

$$Accuracy = 1 - Error\ Rate \qquad \text{Eq. 4}$$

Likewise, OBO accuracy per trial was one if $abs(\ \#Actual\ Reps - \#Predicted\ Reps) \leq 1$ and zero otherwise. Thus, the mean OBO Accuracy was the total number of ones divided by the number of trials.

**Table 7.** Evaluation Results of the Custom Dataset

| Exercise | # Actual Reps | # Predicted Reps | Mean Accuracy | Mean OBO Accuracy | Absolute Accuracy |
|---|---|---|---|---|---|
| *Squats* | 174 | 172 | 88.24% | 85.19% | 98.85% |
| *Push-ups* | 248 | 192 | 65.72% | 60.71% | 77.42% |
| *Pull-ups* | 248 | 203 | 85.96% | 86.21% | 81.85% |
| TOTAL | 670 | 567 | - | - | |
| AVERAGE | - | - | 79.97% | 77.37% | 86.04% |

## 5.2 Real-life evaluation

To test the proposed AERC in real-life scenarios [42], two participants (one male and one female) carried out a series of tests via the app, either at the gym or in a natural environment conducted on various days. Overall, the participants performed a total of 450 repetitions over 17 days and used Pūioio to count the exercises automatically. Tables 8 to 10 provide detailed results of squats, push-ups, and pull-up experiments, respectively, while Table 11 summarizes these results.

**Table 8.** Squat Results

| Subjects | Number of Reps App Counted | Number of Reps Completed | Diff | ACC | OBO | Notes |
|---|---|---|---|---|---|---|
| *Female Subject* | 10 | 11 | 1 | 0.9 | 1 | Did not squat deep enough for one rep |
| *Female Subject* | 10 | 10 | 0 | 1 | 1 | |
| *Female Subject* | 10 | 10 | 0 | 1 | 1 | |
| *Female Subject* | 10 | 10 | 0 | 1 | 1 | |
| *Female Subject* | 10 | 10 | 0 | 1 | 1 | |
| *Male Subject* | 15 | 15 | 0 | 1 | 1 | |
| *Female Subject* | 10 | 10 | 0 | 1 | 1 | |
| *Female Subject* | 10 | 10 | 0 | 1 | 1 | |
| *Male Subject* | 10 | 10 | 0 | 1 | 1 | |
| *Male Subject* | 9 | 9 | 0 | 1 | 1 | Done at a low angle |
| *Male Subject* | 10 | 10 | 0 | 1 | 1 | Done at a low angle |
| *Male Subject* | 8 | 8 | 0 | 1 | 1 | Done at a low angle |
| *Male Subject* | 8 | 8 | 0 | 1 | 1 | Done at a low angle |
| *Female Subject* | 10 | 10 | 0 | 1 | 1 | Done at a low angle |
| *Male Subject* | 9 | 9 | 0 | 1 | 1 | |

**Table 9.** Push-ups Results

| Subject | Number of Reps App Counted | Number of Reps Completed | Diff | ACC | OBO | Notes |
|---|---|---|---|---|---|---|
| *Female Subject* | 10 | 11 | 1 | 0.9 | 1 | Didn't do the first push-up deep enough |
| *Female Subject* | 10 | 10 | 0 | 1 | 1 | Hands were positioned wider than shoulders |
| *Female Subject* | 10 | 10 | 0 | 1 | 1 | Hands were positioned wider than shoulders |
| *Male Subject* | 10 | 10 | 0 | 1 | 1 | |
| *Female Subject* | 10 | 10 | 0 | 1 | 1 | |
| *Female Subject* | 10 | 10 | 0 | 1 | 1 | |
| *Female Subject* | 10 | 10 | 0 | 1 | 1 | |
| *Female Subject* | 10 | 10 | 0 | 1 | 1 | |
| *Male Subject* | 10 | 10 | 0 | 1 | 1 | |
| *Male Subject* | 10 | 10 | 0 | 1 | 1 | |
| *Male Subject* | 15 | 15 | 0 | 1 | 1 | |
| *Male Subject* | 15 | 15 | 0 | 1 | 1 | |
| *Male Subject* | 10 | 10 | 0 | 1 | 1 | |
| *Male Subject* | 9 | 9 | 0 | 1 | 1 | |

**Table 10.** Pull-up Results

| Subjects | Number of Reps App Counted | Number of Reps Completed | Diff | ACC | OBO | Notes |
|---|---|---|---|---|---|---|
| *Female Subject* | 10 | 10 | 0 | 1 | 1 | |
| *Female Subject* | 10 | 10 | 0 | 1 | 1 | |
| *Female Subject* | 10 | 10 | 0 | 1 | 1 | |
| *Female Subject* | 10 | 10 | 0 | 1 | 1 | |
| *Female Subject* | 10 | 10 | 0 | 1 | 1 | |
| *Female Subject* | 10 | 10 | 0 | 1 | 1 | |
| *Female Subject* | 10 | 10 | 0 | 1 | 1 | |
| *Male Subject* | 10 | 10 | 0 | 1 | 1 | |
| *Male Subject* | 10 | 10 | 0 | 1 | 1 | Done at different angles |
| *Male Subject* | 10 | 10 | 0 | 1 | 1 | |
| *Male Subject* | 10 | 10 | 0 | 1 | 1 | |
| *Male Subject* | 10 | 7 | 3 | 0.57 | 0 | Wore black long sleeve resulting in the system struggling to detect the pose |
| *Male Subject* | 10 | 10 | 0 | 1 | 1 | |
| *Male Subject* | 20 | 20 | 0 | 1 | 1 | Phone facing sun |
| *Male Subject* | 3 | 3 | 0 | 1 | 1 | |

**Table 11.** Summary of Results from Testing Pūioio

| Exercise | # Actual Reps | # Predicted Reps | Mean Accuracy | Mean OBO Accuracy | Absolute Accuracy |
|---|---|---|---|---|---|
| *Squats* | 150 | 149 | 99.33% | 100% | 99.33% |
| *Push Ups* | 150 | 149 | 99.28% | 100% | 99.33% |
| *Pull-ups* | 150 | 153 | 97.13% | 93.33% | 98% |
| TOTAL | 450 | 451 | - | - | |
| AVERAGE | - | - | 98.58% | 97.77% | 98.89% |

## 6 DISCUSSION AND CONCLUSION

In this paper, a novel exercise repetition counting pipeline is proposed. Furthermore, a cross-platform mobile application has been developed to run the pipeline on-device and offline. Due to limited open-source datasets, a video dataset was collated, consisting of 90 videos for three exercises. Our model counts valid repetitions with over 85% accuracy for two of the three exercises when testing the dataset. On the other hand, we consistently achieved over 98% accuracy for all three exercises when conducting real-world tests. These desirable results confirm that the final application is beneficial for tracking repetitions during a workout using only a smartphone camera and without requiring inference to be done remotely on a server. The application, Pūioio, is available for download via the Google Play Store and Apple App Store.

Our evaluation of the system using the custom dataset of 30 videos per exercise (total of 670 repetitions) achieved an absolute accuracy of 86.04%. However, the system achieved an absolute accuracy of 98.89% when it was used to count 450 repetitions performed by the test subjects in real-life testing. The lower performance obtained from evaluating the system by the custom dataset of pre-recorded videos can be attributed to the videos selected for the dataset, as they included:

1. Individuals whose bodies were not fully visible in the frame.
2. Individuals who did not perform complete exercises as defined by consulted professionals. That is, repetitions were not completed to the model's standard of counting a repetition, as the user did not reach the predefined threshold(s).
3. Individuals whose bodies were positioned at an acute angle in the frame.

During testing with pre-recorded videos, any video sample failing to achieve an accurate count due to the subject's body not being entirely captured was not disregarded to simulate more realistic conditions that resulted in the proposed AERC's miscount. Furthermore, those video samples of users who did not perform the exercises with adequate depth or range were insignificant as they technically did not execute a correct exercise repetition, yet such situations were still considered miscounts and reflected in the results reported. Lastly, BlazePose, which was used for pose estimation, could not determine the pose of an individual whose face was not visible during the dataset experiments, resulting in miscounts for people angled away from the camera due to the pose estimation not being able to predict key points correctly. The app rectifies these limitations as the system conveys to the user when they have reached each phase to ensure they complete a proper repetition in real-world usage and pauses counting if it fails to find the user.

The real-life testing results were achieved in real-time without any discernible lag that otherwise would have negatively affected the user experience. Our approach was designed to cease counting when the subject's body cannot be confidently detected in the frame after 1.5 seconds. During the real-life experiments, the app was consistently accurate and easy to use when testing in a gym environment. It had no trouble with people walking in the background, nor did it have issues with different angles if the test subject had their full body in the frame.

The current architecture could be extended using a basic, mobile-optimized convolutional neural network in an ensemble arrangement with pose detection, thresholding, and optical flow. If this neural network can learn key features distinguishing a subject and exercise moments, it would greatly increase the architecture's robustness to challenges such as occlusion and lack of contrast in clothing, with which the architecture currently struggles. Future work could also explore using a sequential model to take advantage of the temporal dimension of the data rather than only operating on an independent, frame-by-frame basis. Though this frame-by-frame approach was successful, information is lost in the relation between frames; thresholding misses this entirely, and optical flow only begins to capture it in its two-frame approach. Intuitively, this lost information reduces the architecture's ability to adapt to more complex exercises and environments, such as when body parts are occluded during repetitions.

**Statements and Declarations**

Competing Interests: the authors declare no financial or competing interests.

Data Availability: The datasets generated during and/or analysed during the current study are available from the corresponding author on reasonable request.